\documentclass[12pt,onecolumn]{pasj00}

\usepackage{graphicx}
\newcommand{\Vec}[1]{\mbox{\boldmath $#1$}}

\SetRunningHead{Maki \& Susa}{Dissipation of Magnetic Flux in Primordial
Star Formation}

\title{Dissipation of Magnetic Flux in Primordial Star Formation:\\
       From Run-away Phase to Mass Accretion Phase}
\author{Hideki Maki}
\affil{Department of Physics, Rikkyo University,
       Nishi-Ikebukuro, Tokyo 171-8501}
\email{hide-mk@jcom.home.ne.jp}

\author{Hajime Susa}
\affil{Department of Physics, Konan University,
       Okamoto, Kobe 658-8501}
\email{susa@center.konan-u.ac.jp}

\Received{$\langle$reception date$\rangle$}
\Accepted{$\langle$acception date$\rangle$}
\Published{$\langle$publication date$\rangle$}

\begin{document}
\maketitle
\begin{abstract}
We investigate the dissipation of magnetic flux
in primordial star-forming clouds throughout their collapse
 including the run-away collapse phase
as well as the accretion phase.
We solve the energy equation and
the non-equilibrium chemical reactions in the
collapsing gas, in order to obtain the radial distribution of the
 ionized fraction during the collapse.
As a result, we find the ionized fraction is high enough for the
 magnetic field to couple with the gas throughout the evolution of the
 cloud.
This result suggests that the jet formation from protostars as well
 as the activation of magneto-rotational instability in the accretion disk
 are enabled in the presence of the cosmological seed magnetic flux
 proposed by \citet{langer03}.
\end{abstract}


\section{INTRODUCTION}
\label{intro}
Typical mass or initial mass function of population III stars are
fundamental parameters that have great impacts on subsequent structure
formation of the universe. Those stars are expected to be as massive as
$100-1000\MO $, 
which ionize/dissociate the surrounding media. As a result, star formation
activities in the neighbourhood of the stars are highly regulated
(e.g. \citet{Susa07} and the references are therein).

There are two chief reasons that the population III stars are expected
to be very massive. First one comes from numerical
studies by several authors
\citep{nakamura99,bromm99,abel00,nakamura01,bromm02}.
Those studies reveal that the prestellar cores formed as fragments of
primordial gas are as massive as $\sim 10^3 \MO-10^4 \MO$.
Second ground is the very high accretion rates of those
stars, which is as high as $10^{-3}-10^{-2}\MO{\rm yr}^{-1}$. These two
facts are the direct consequence of relatively high temperature
($\sim 1000$K) of primordial gas because of inefficient cooling by
H$_2$.

On the other hand,
abundance pattern of the hyper metal-poor stars seems to be more
consistent with that of ``faint'' supernovae as remnants of less
massive ($\sim 25 \MO$) population III
stars\citep{Christlieb04,Frebel05,Iwamoto05}. 
Recent theoretical studies
suggest the possibilities for the formation of such less massive
population III stars. 
If the primordial gas was once ionized, enhanced fraction of H$_2$
causes more efficient cooling and HD formation. Since HD molecules have
lower excitation energy than that of H$_2$, gas can be cooled below
100K. As a result, the mass of the fragments could be smaller than
$100\MO$\citep{oshea05,JB06}. \citet{oshea07} also demonstrates that such low
mass population III stars cloud be formed directly from the cosmological
density fluctuations.

Those new ideas are quite interesting and promising, 
however, we still do not know the
actual mass of population III stars formed from $\sim 10^3 \MO -10^4 \MO$
prestellar cores, which are commonly found in cosmological simulations. 
Since they are only $10^4-10^6 {\rm cm^{-3}}$, we have
to follow the subsequent evolution of the collapsing cloud.
Further evolution of prestellar cores is first investigated by
\citet{omukai98}, and they found that the collapse proceeds in a run-away
fashion and converges to Larson-Penston type similarity solution
\citep{Larson69,Penston69,suto88} with polytropic index 
$\gamma\simeq 1.09$.
They also find the mass accretion rate is very
large compared to the present-day forming stars, although spherical
symmetry is assumed in their radiation hydrodynamic simulations. Recently,
\citet{Yoshida06} have performed cosmological simulations in which the
run-away collapsing core is traced up to a very dense regime ($n_{\rm H}
\gg 10^{10}{\rm cm^{-3}}$), taking the Sobolev type line transfer
approximations. They also find basically consistent results in 3D
cosmological simulations with previous 1D results by \citet{omukai98}.

However, most of the mass of protostar is accumulated in the accretion
phase, which has not been traced especially in multi-dimensional
simulations. Recent numerical simulations performed by \citet{saigo05}
show that disks would be formed at the center of collapsing
primordial clouds, 
which might result in the disks surrounding the protostars, or
binaries. Thus, the actual mass of a population III star should depends on the
mechanism of the angular momentum transport in the accretion disk.
There are a few possibilities of the angular momentum transport mechanism
such as gravitational torque by the nonaxisymmetric structures in the
accretion disk, the interaction among the 
fragments of the disk \citep{stone00,bodenheimer00}, 
and the turbulent viscosity
triggered by Magneto-Rotational Instability (MRI)
\citep{hawley92,sano98,sano01}.
As for MRI induced turbulence, the strength of magnetic field in 
the disk is the key quantity to activate the instability\citep{TM04,TB04}.
It is also worth noting that 
recent MHD simulations by \citet{machida06} suggests the
possibility of bipolar jets from proto-population III stars, which also
could suppress the mass accretion onto the central core. The formation
of jets also requires the presence of some level of magnetic field.
Therefore, it is quite important to assess the magnetic field
strength brought into the accretion disk from initial weak cosmological
seed field (e.g. \citet{langer03}).

\citet{MS04} investigated this issue by solving detailed chemical
reaction rate equations coupled with energy equation in run-away collapsing
core. They found that magnetic field is always frozen to the collapsing
core in case the strength of initial magnetic field is smaller than
$10^{-5}(n_{\rm H}/10^3~{\rm cm^{-3}})^{0.55}~{\rm G}$. This is
comparable to the maximal strength of magnetic field which allow the
clouds to collapse. Therefore they conclude magnetic field is always
frozen to the {\it collapsing} cloud in the run-away collapsing core.
They also evaluate the minimal strength of magnetic field which is
required to activate MRI assuming magnetic field is frozen to the gas
not only in the run-away phase, but also in the accretion phase.

This assumption is based upon the argument that the temperature of the
accretion flow would rise faster than that of the core, because of the
shock heating. As a result, the ionization degree is expected to be
higher than those in run-away phase, which guarantee the gas to be frozen
to the magnetic field lines.
However, such heating is only important in the final
phase of accretion, where the flow settle onto the accretion disk. 
In order to investigate the coupling of matter and magnetic field in the
accretion phase, we need to calculate the actual thermal and chemical
evolution of accretion flow.
Aside from the issue on ionization degree, 
the ambipolar diffusion velocity is proportional to the square
inverse of density ($v_{\rm B}^{\rm amb}\propto \rho^{-2}$). 
As a result, 
the diffusion velocity increases rapidly in the mass accretion phase,
since the density at a fixed radius decreases as the accretion proceeds.
Thus, we need a detailed treatment to test the coupling between the gas
and magnetic field in accretion phase.

In this paper, we investigate the dissipation of magnetic field in
collapsing primordial gas cloud from run-away phase to accretion phase,
by solving detailed thermal and chemical rate equations.
In the next section, we describe the formulations employed. 
In \S\ref{results}, results of our calculations are shown. The
formation of jets and activation of MRI is discussed
in \S\ref{discussion}. Final section is devoted to summary.

\section{METHOD OF CALCULATIONS}
\label{models}
In order to evaluate the coupling of the gas and the magnetic field, we need
to assess the amount of ions and electrons in collapsing primordial
cloud. The collapse of the cloud is expected to proceed in run-away
fashion (run-away phase) in the beginning,  followed by the mass
accretion phase after the formation of rotationally supported disk at
the center. In this paper, we follow the chemical and thermal evolution
of the materials to form star, throughout the two phases. We assume
spherically symmetric collapse of the progenitor gas, although we expect
the formation of rotationally supported disk in the very dense regime
($n_{\rm H} \gg 10^{10} {\rm cm^{-3}}$). It is also assumed that the
magnetic field is so weak that the dynamics of the collapsing gas 
is not affected by the magnetic force.

\subsection{Run-away Collapse Phase}
\label{runaway}
The run-away collapse phase is traced by solving 1-dimensional hydrodynamics.
The equation of continuity is 
\begin{equation}
\frac{dm}{dr}=4\pi r^2\rho,
\end{equation}
whereas the equation of motion is
\begin{equation}
\frac{Du}{Dt}=-4\pi r^2\frac{\partial p}{\partial m}-\frac{Gm}{r^2},
\end{equation}
where $m$ is the mass within radius $r$, $\rho$ is the density of the cloud,
$u$ is the velocity and $p$ is the pressure.
Since the equation of state of collapsing primordial gas in run-away phase
is known to be approximated by polytrope with $\gamma=1.09$ \citep{omukai98}, we employ following equation:
\begin{equation}
p=K\rho^\gamma, \quad \gamma=1.09
\end{equation}
We solve above set of equations by spherically symmetric Lagrangian
hydrodynamics code developed by ourselves, following the Piecewise
Parabolic Method (PPM) described in \citet{CW84}.

\subsection{Mass Accretion Phase}
\label{accretionphase}
\begin{figure}[tbh]
\begin{center}
\FigureFile(13cm,10cm){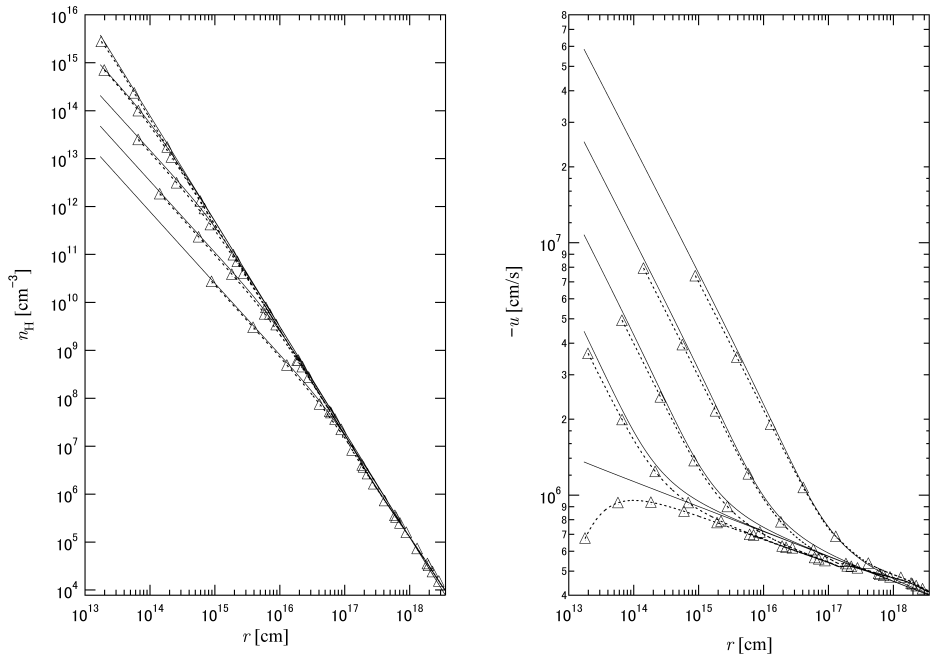}
\caption{
Time evolution of density/velocity profiles in accretion
 flow are shown. Dotted lines and marks represent the result with free-fall
 approximation,
 whereas the solid lines denote the results from the similarity
 solution in accretion phase with $\gamma=1.09$.
}
\label{fig:free-fall v.s. polytrope}
\end{center}
\end{figure}

The dynamics of accretion phase could be approximated by simple free-fall,
since the flow velocity is supersonic.
Basic equations of free-falling accretion flow consists of 
the equation of continuity and the equation of motion:
\begin{equation}
4\pi\rho(t_0,r_0)r_0^2dr_0=4\pi\rho(t,r)r^2(t)dr,
\label{eq:continuity at accretion phase}
\end{equation}
\begin{equation}
\frac{d^2r\left(t\right)}{dt^2} = -\frac{GM_0}{r^2(t)},
\label{eq:eq. of motion in accretion phase}
\end{equation}
where $r$, $t$ and $t_0$ represent 
the position of fluid element, time, and the time when the mass
accretion starts, respectively.
$M_0$ denotes the mass within $r_0$ at $t=t_0$:
\begin{equation}
M_0 = \int_0^{r_0}4\pi r'_0\rho_0dr'_0.
\end{equation}
The initial radius $r_0$ is defined as $r_0 \equiv r(t_0)$.
Therefore, the the solution of above equation of motion is given as 
$r=r(t;t_0,r_0)$.
The velocity of the fluid element is given by
\begin{equation}
u(t;t_0,r_0) = -\sqrt{2\mathcal{E}_0+2\frac{GM_0}{r(t)}},
\end{equation}
where $\mathcal{E}_0$ is the total energy defined as 
$\mathcal{E}_0\equiv u^2(t_0)/2-GM_0/r_0$.
Using equation (\ref{eq:continuity at accretion phase}),
the density at $(t,r)$ is given as 
\begin{equation}
\rho(t,r) = \frac{r_0^2}{r^2}
  \frac{\rho_0}{\left(\partial r/\partial r_0\right)_t}\label{eq:density in accretion phase}.
\end{equation}
$(\partial r/\partial r_0)_t$ is the partial differentiation by $r_0$
keeping $t$ fixed. An explicit expression of this term is
given in appendix \ref{appendix_a}.
In order to clarify the validity of the free-fall approximation, it is compared with the similarity solution
with $\gamma=1.09$ \citep{suto88} in accretion phase. The initial
condition of the accretion flow is set as the final state of the
run-away collapse with $\gamma=1.09$. Figure \ref{fig:free-fall v.s. polytrope} illustrates the density and velocity distributions of several
snapshots. It is obvious that the free-fall approximation can describe
the matter distribution in accretion phase very well for polytropic gas.
In other words, the thermal
evolution has little effects on the dynamics of the accretion flow. Thus,
the dynamics can be approximated by free-fall formula, irrespective of
the internal energy equation.

In the accretion flow, the effective polytropic index is basically
unknown. 
Thus, we have to solve the following energy equation for each fluid
element:
\begin{equation}
\frac{d\varepsilon}{dt} =
  -p\frac{d}{dt}\left(\frac{1}{\rho}\right)-\mathcal{L}^{\rm(net)}.
\end{equation}
In this expression, $p$ denotes the pressure, which is related to the internal energy
$\varepsilon$ by the equation of state: 
\begin{equation}
p = (\gamma_{\rm ad}-1)\rho\varepsilon,
\end{equation}
where $\gamma_{\rm ad}$ is the adiabatic exponent, $\mathcal{L}^{\rm
(net)}$ denotes the net energy loss rate per unit mass. 

We  take into account the rovibrational line cooling by $\rm H_2$($\mathcal{L}_{\rm line}$),
the continuum radiation from the gas ($\mathcal{L}_{\rm cont}$),
and cooling and heating associated with chemical reactions
($\mathcal{L}_{\rm diss},~\mathcal{G}_{\rm H^-},~\mathcal{G}_{\rm
H^+_2}$ and $\mathcal{G}_{\rm 3body}$). $\mathcal{L}^{\rm (net)}$ is the
sum of all these contributions.

The cooling rate by the rovibrational transition of hydrogen molecules
are assessed by the fitting formula given by \citet{galli98} in case the
lines are optically thin. In the optically thick regime, we employ the
escape probability formalism, as described in equations
(12)-(15) in \citet{omukai00}, except that we use actual velocity gradient
in our calculations.


The cooling rate owing to the continuum radiation is given by
\begin{equation}
\mathcal{L}_{\rm cont} = 4\sigma \kappa_{\rm gas} T^4
\end{equation}
where $\sigma$ denotes the Stefan-Boltzmann constant, 
$\kappa_{\rm gas}$ is the Planck mean opacity of gas \citep{lenzuni91}, which includes  bound-free absorption,
free-free absorption, photodissociation,
Rayleigh scattering, and collision-induced absorption.
This formula is valid in case the accreting gas is optically thin for
continuum radiation. We confirmed that the optical depth is smaller than
unity throughout the accretion phase.

Cooling due to the latent heat of H$_2$ dissociation is given by
\begin{equation}
\mathcal{L}_{\rm diss}=4.48\frac{n_{\rm H}}{\rho}
  \left(\frac{dy_{\rm H_2}}{dt}\right)_- {\rm eV\;s^{-1}\;g^{-1}},
\end{equation}
where $n_{\rm H}$ is the number density of hydrogen nuclei,
and $(dy_{\rm H_2}/dt)_-$ is the dissociation rate of $\rm H_2$.

On the other hand, the gas is heated when 
a hydrogen molecule is formed.
The heating rate per unit mass is given by
\begin{eqnarray}
& \mathcal{G}_{\rm H^-} = \frac{3.53}{\rho}
   \left(\frac{dy_{\rm H_2}}{dt}\right)_{\rm H^-}
    \frac{n_{\rm H}}{1+n_{\rm cr}/n_{\rm H}}{\rm\;eV\;s^{-1}\;g^{-1}}, \\
& \mathcal{G}_{\rm H^+_2} = \frac{1.83}{\rho}
   \left(\frac{dy_{\rm H_2}}{dt}\right)_{\rm H^+_2}
    \frac{n_{\rm H}}{1+n_{\rm cr}/n_{\rm H}}{\rm\;eV\;s^{-1}\;g^{-1}}, \\
& \mathcal{G}_{\rm 3body} = \frac{4.48}{\rho}
   \left(\frac{dy_{\rm H_2}}{dt}\right)_{\rm 3body}
    \frac{n_{\rm H}}{1+n_{\rm cr}/n_{\rm H}}{\rm\;eV\;s^{-1}\;g^{-1}},
\end{eqnarray}
where $(dy_{\rm H_2}/dt)_{\rm H^-}$, $(dy_{\rm H_2}/dt)_{\rm H^+_2}$
and $(dy_{\rm H_2}/dt)_{\rm 3body}$ are
the formation rates of $\rm H_2$  by H$^-$ process, H$_2^+$ process and
three body reactions \citep{HM79}.
$n_{\rm cr}$ is the critical density defined as
\begin{eqnarray}
n_{\rm cr} = \frac{10^6}{T^{1/2}}
 \Biggl\{1.6y_{\rm H}\exp\left[-\left(\frac{400}{T}\right)^2\right]
  + 1.4y_{\rm H}\exp\left[-\left(\frac{1.2\times10^4}{T+1200}\right)\right]
 \Biggr\}^{-1}\;{\rm cm^{-3}}.
\end{eqnarray}
\subsection{Chemical Reactions}
Since the dissipation of the magnetic flux,
as well as the temperature of the gas,
strongly depends on the chemical abundances,
we have to solve the non-equilibrium chemical reaction rate equations
coupled with equation of motion and energy equation described in previous
subsections \ref{runaway} and \ref{accretionphase}.

The evolution of the fraction of species $i$ is followed by 
solving the equations,
\begin{eqnarray}
\frac{dy_i}{dt} &=& \sum_{l=1}^{24}\sum_{m=1}^{24}n_{\rm H}k_{lm}y_ly_m
  + \sum_{l=1}^{24}\sum_{m=1}^{24}\sum_{n=1}^{24}n_{\rm H}k_{lmn}y_ly_my_n,
  \nonumber \\
  & &\qquad\qquad\qquad\qquad\qquad\quad(i=1,2,3,\cdots,24),
\end{eqnarray} 
where $y_i\equiv n_i/n_{\rm H}$ is the fraction of species $i$,
$k_{lm}\;{\rm [cm^3\;s^{-1}]}$ and
$k_{lmn}\;{\rm [cm^6\;s^{-1}]}$ are
the reaction rate coefficients
with respect to two-body processes and three-body processes, respectively.
In our calculation, we include 24 species:
$\rm e^-$, $\rm H^+$, H, $\rm H^-$, $\rm H_2$, $\rm H_2^+$,
$\rm H_3^+$, D, $\rm D^+$, $\rm D^-$, HD, $\rm HD^+$, $\rm H_2D^+$,
He, $\rm He^+$, $\rm He^{++}$, $\rm HeH^+$, Li, $\rm Li^+$,
$\rm Li^{++}$, $\rm Li^{3+}$, $\rm Li^-$, LiH, and $\rm LiH^+$.
We employ the latest reaction rate coefficients in the
following papers, \citet{galli98}, \citet{omukai00},
\citet{stancil98}, \citet{flower02} and \citet{lepp02}.
As for the radiative recombination, we use the rate coefficients based on
\citet{spitzer78}. These reactions are the same as our previous paper\citep{MS04}.
Here we stress the importance to include above rare elements such as Li,
because the coupling of magnetic field with gas can be attained by
very low fractional abundance of electrons and ions \citep{MS04}.

\subsection{Drift velocity of magnetic field}
Magnetic field is dissipated from star-forming gas via ohmic loss 
and ambipolar diffusion.
We assess the drift velocity $v_{{\rm B}x}$ of the field lines due to
these two processes, which is compared to the accretion velocity of gas. 
We evaluate the drift velocity following the formulation by
\citet{nakano86}.
There are two important quantities which characterize
these diffusion processes. They are $\tau_\nu$ and $\omega_\nu$ which denote
the viscous damping time of the relative velocity of charged particle $\nu$
to the neutral particles,  and the cyclotron frequency of
the charged particle $\nu$, respectively. 
Then,  $\tau_\nu$ is expressed as
\begin{equation}
	\tau_\nu = \frac{\rho_\nu}{\mu_{\rm \nu n}
				n_\nu n_{\rm n}\langle\sigma v\rangle_{\rm \nu n}},
\end{equation}
where $\mu_{\rm \nu n}$ is the reduced mass, $n_\nu$, $n_{\rm n}$,
and $\rho_\nu$ are, the mean number density for
the charged particle $\nu$, the neutral particle ${\rm n}$, and the mass
density of charged particle $\nu$, respectively. The averaged  momentum-transfer rate coefficient for a
particle $\nu$ colliding with a neutral particle is expressed by  $\langle \sigma v \rangle_{\rm \nu n}$ .
We use the empirical formulae for the momentum-transfer rate coefficients
\citep{kamaya00,sano00}.

According to \citet{nakano86}, the drift velocity
is given by
\begin{equation}
	v_{{\rm B}x} = \frac{A_1}{A}\frac{1}{c}(\Vec{j}\times\Vec{B})_x,
\label{eq:diffusion_velocity}
\end{equation}
where
\begin{eqnarray}
	A = A_1^2 + A_2^2, \label{eq:A}\\
	A_1 = \sum_\nu \frac{\rho_\nu \tau_\nu \omega_\nu^2}
		{1 + \tau_\nu^2 \omega_\nu^2}, \label{eq:A1}\\
	A_2 = \sum_\nu \frac{\rho_\nu \omega_\nu}
		{1 + \tau_\nu^2 \omega_\nu^2}, \label{eq:A2}
\end{eqnarray}
$\Vec{B}$ is the mean magnetic field in the primordial cloud,
the suffix $x$ means $x$ direction component in local Cartesian
coordinates where the z direction is taken as the direction of
$\Vec{B}$.
 We replace
$(1/c)(\Vec{j} \times \Vec{B})_x$ in equation
(\ref{eq:diffusion_velocity}) with the mean magnetic force
$B^2/4 \pi r$, where $B$ is the mean field strength in the cloud,
$r$ is the radius of the cloud at which we assess the drift velocity.

\subsection{Initial Conditions}
We consider primordial star-forming gas clouds that formed in the
mini-halos with $\sim 10^6\;\MO$.
We set uniform and spherical gas cloud with $n_{\rm H}=10^3\;{\rm
cm^{-3}}$, $T_0=250\;{\rm K}$, and $M=10^4\;\MO$.
Such clouds are commonly found 3D cosmological simulations
\citep{bromm99,abel00,abel02,bromm02,Yoshida03,oshea06}.
We also use the cosmological abundance given by \citet{galli98},
as the initial fraction of the chemical compositions.
The initial magnetic field strength is expected to be very weak,
however, its magnitude is still under discussion. Thus, we regard the
field strength as a free parameter of the calculations.
 
\subsection{Switching from Run-away Phase to Accretion Phase}
The run-away phase and the subsequent accretion phase have to be treated
in different formulation from each other as described in subsections
\ref{runaway} and \ref{accretionphase}. Thus, we have to switch the
scheme from the method in run-away phase to that in
accretion phase. In the light of physical arguments, 
two phases should be switched when the central run-away
collapse is stopped due to thermal pressure or centrifugal force.
In our calculation, we switch the scheme when the
sonic point of the accreting flow 
go inside a certain small radius. We set this radius to be
$r_{\rm sw}=700\;\RO$, 
which is comparable to the disk size when the accreted
mass is comparable to $\sim 1\MO$\citep{TM04}. It is also worth noting
that the central density of the core at switching is $\sim 10^{16}~{\rm cm^{-3}}$.

\begin{figure}[htb]
\begin{center}
\FigureFile(12cm,10cm){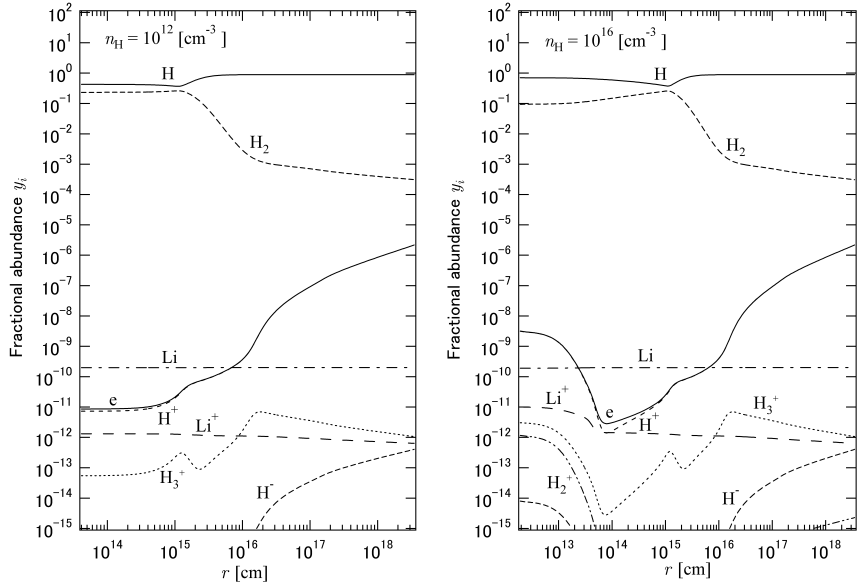}
\caption{
The distribution of the main species, $\rm e^-$,
$\rm H^+$, H, $\rm H_2$, $\rm H_2^+$, $\rm H_3^+$,
Li, and $\rm Li^+$ are plotted.
The vertical axis denotes the fractional abundance $y_i$
of the above species, and the horizontal axis
denotes the radius from the core center.
Left panel represents the snapshot when the central density satisfies $n_{\rm H,c}=10^{12}\;\rm cm^{-3}$,
whereas right panel shows the results at $n_{\rm H,c} = 10^{16}\;\rm cm^{-3}$.
}
\label{fig:chemi_runaway}
\end{center}
\end{figure}
\begin{figure}[bht]
\begin{center}
\FigureFile(12cm,7cm){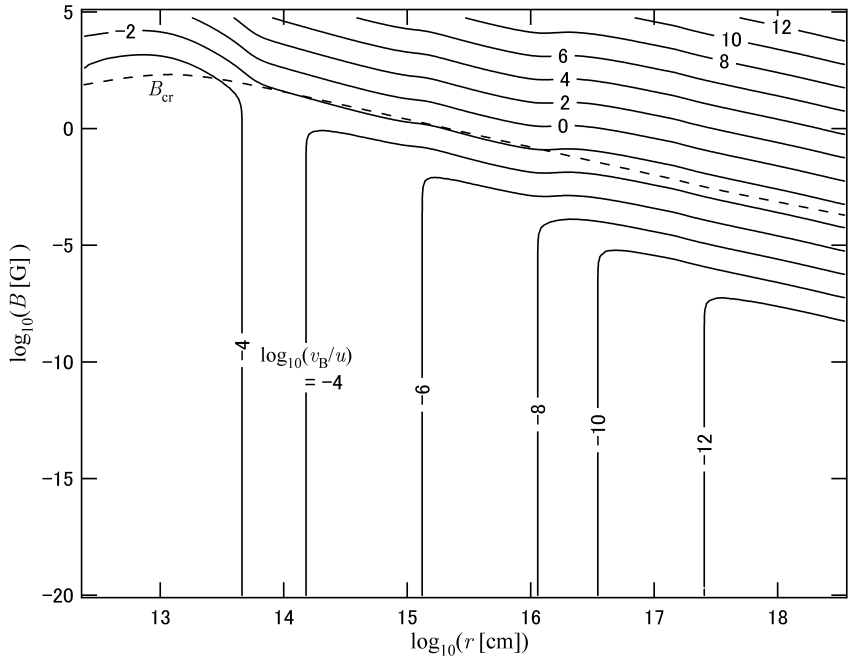}
\caption{Drift velocity $v_{{\rm B}x}$ as a function
of the radius $r$ and the field strength $B$ at 
the central density .
Contour map of $v_{{\rm B}x}/u$ is shown on $r-B$ plane for $n_{\rm H,c} = 10^{16}\;{\rm cm^{-3}}$
Solid curves represent the constant loci along which $\log(v_{{\rm
 B}x}/u)$ equal to the values labeled on the curve.
The dashed curve represents the critical field strength $B_{\rm cr}$
given by the equation (\ref{eq:critical field}).}
\label{fig:rdiffu_runaway}
\end{center}
\end{figure}

\section{RESULTS}
\label{results}
\subsection{Runaway Collapse Phase}
\subsubsection{The distribution of the chemical abundances}
Two snapshots of $y_i$ distribution during the run-away
collapse phase are shown in Figure \ref{fig:chemi_runaway}.
The left panel shows the distribution
at the time when the central density $n_{\rm H,c}$
reaches $10^{12}\;{\rm cm^{-3}}$,
whereas the right panel represents the distribution
at $n_{\rm H,c} = 10^{16}\;{\rm cm^{-3}}$.
The horizontal axis denotes the radius from the center
of the primordial gas cloud $r\;[\rm cm]$, and
the vertical axis is the fractional abundances $y_i$
of each species.
Since the central density increases as the collapse proceeds,
right panel represents the later epoch than that of the left.

Roughly speaking, in both of the panels, fraction of electrons ($y_{\rm
e}$) in the inner
region are smaller than that in the outer part. These results could be
simply understood since the recombination process in the inner dense
region proceeds faster than in the outer envelope. We also find the opposite
behavior in the inner most region of right panel,
at which $y_{\rm e}$ increases as $r$ decreases, because of the
collisional ionization. It is also worth noting that $\rm Li$ becomes the
main provider of electrons around $r\sim 10^{14}{\rm cm}$ at $n_{\rm H,c} = 10^{16}\;{\rm cm^{-3}}$. Because of the combined effects of these two
(collisional ionization \& presence of Li), $y_{\rm e}$ never gets lower
than $10^{-11}$ as far as we consider $n_{\rm H,c} < 10^{16}\;{\rm
cm^{-3}}$ in run-away phase.  
\subsubsection{Drift velocities in run-away collapse phase}
\label{drift_runaway}
In Figure \ref{fig:rdiffu_runaway},
we show the ratio of the drift velocity $v_{{\rm B}x}$ to the
infall velocity $u$ 
when $n_{\rm H,c} =10^{16}\;{\rm cm^{-3}}$ is satisfied.

The vertical axis shows the magnetic field strength, whereas the
horizontal axis shows the distance from the cloud center. 
The solid curves in Figure \ref{fig:rdiffu_runaway}
show the contours of $\log_{10}(v_{{\rm B}x}/u)$.
The magnetic fields are dissipated in 
the region where $\log_{10}(v_{{\rm B}x}/u)>0$,
in contrast, the fields are frozen to the gas
in the region where $\log_{10}(v_{{\rm B}x}/u)<0$.
The dashed curve in Figure \ref{fig:rdiffu_runaway}
is the critical field strength $B_{\rm cr}$
that is defined by the equation
\begin{equation}
\frac{B_{\rm cr}^2}{4\pi r} = \frac{GM(r)\rho(r)}{r^2}.
\label{eq:critical field}
\end{equation}
Note that since we are interested in the collapsing gas cloud,
the magnetic force needs to be weaker than the gravitational force.
Hence, our calculations are valid in the region
where the field strength in the cloud satisfies $B<B_{\rm cr}$.
We find clearly from Figure \ref{fig:rdiffu_runaway}
that the frozen-in condition $v_{{\rm B}x}/u<1$ is almost
always satisfied as long as $B$ is less than the critical
field strength $B_{\rm cr}$. We also find the basically same results for
other snapshots, i.e. the drift velocity is smaller than the infall
velocity anytime and anywhere if $B<B_{\rm cr}$.

In addition, if $B<B_{\rm cr}$ is satisfied at some initial time and
position $(t_0,r_0)$, this
condition also holds at some later epoch 
$(t,r(t))$. This statement is proved as follows:
Combining the magnetic flux conservation equation
\begin{equation}
2\pi r_0 dr_0 B(t_0,r_0) = 2\pi r(t) dr B(t,r(t))
\end{equation}
and the mass conservation law given in equation (\ref{eq:continuity at
accretion phase}), we obtain
\begin{equation}
B(t,r(t))=B(t_0,r_0) \frac{\rho(t,r(t)) r(t)}{\rho(t_0,r_0)r_0},
\label{eq:B frozen}
\end{equation}
Using equations (\ref{eq:critical field}) and (\ref{eq:B frozen}) we
have
\begin{equation}
\frac{B(t,r(t))}{B_{\rm cr}(t,r(t))}=\frac{B(t_0,r_0)}{B_{\rm
 cr}(t_0,r_0)}\left(\frac{\rho(t,r(t)) r^3\left(t\right)}{\rho\left(t_0,r_0\right)r_0^3}\right)^{1/2}
\label{eq:B ratio}
\end{equation}
The second term in right side is always less than or equals to unity 
in the run-away collapsing cloud, 
since $\rho\propto r^{-2}$ in the envelope, and
$\rho\propto r^{-3}$ in the core. 
Thus, if ${B(t_0,r_0)}/{B_{\rm cr}(t_0,r_0)} < 1$ is satisfied, 
$B$ is less than $B_{\rm cr}$ throughout the collapse. 
Remark that we assume the flux conservation as equation (\ref{eq:B
frozen}), which gives the maximal field strength. Therefore, above
inequality ($B<B_{\rm cr}$) also holds even if the magnetic flux is dissipated.

In summary, it is concluded that the magnetic fields
are always frozen to the whole cloud
in the course of run-away collapse phase, if the magnetic force is much
smaller than gravitational force at the beginning of the collapse.

We also remark that the slight increase due to collisional
ionization at $n_{\rm H,c} =10^{16}\;{\rm cm^{-3}}$ implies that $y_{\rm
e}$ will become larger as the central density increases up to $n_{\rm
H,c} > 10^{16}\;{\rm cm^{-3}}$, since the temperature gets higher as the
density increases for $\gamma=1.09$ polytropic gas. As a result, the
magnetic field is expected to be frozen to the gas even for $n_{\rm
H,c} > 10^{16}\;{\rm cm^{-3}}$ in accretion phase.
\begin{figure}[tbh]
\begin{center}
\FigureFile(13cm,7cm){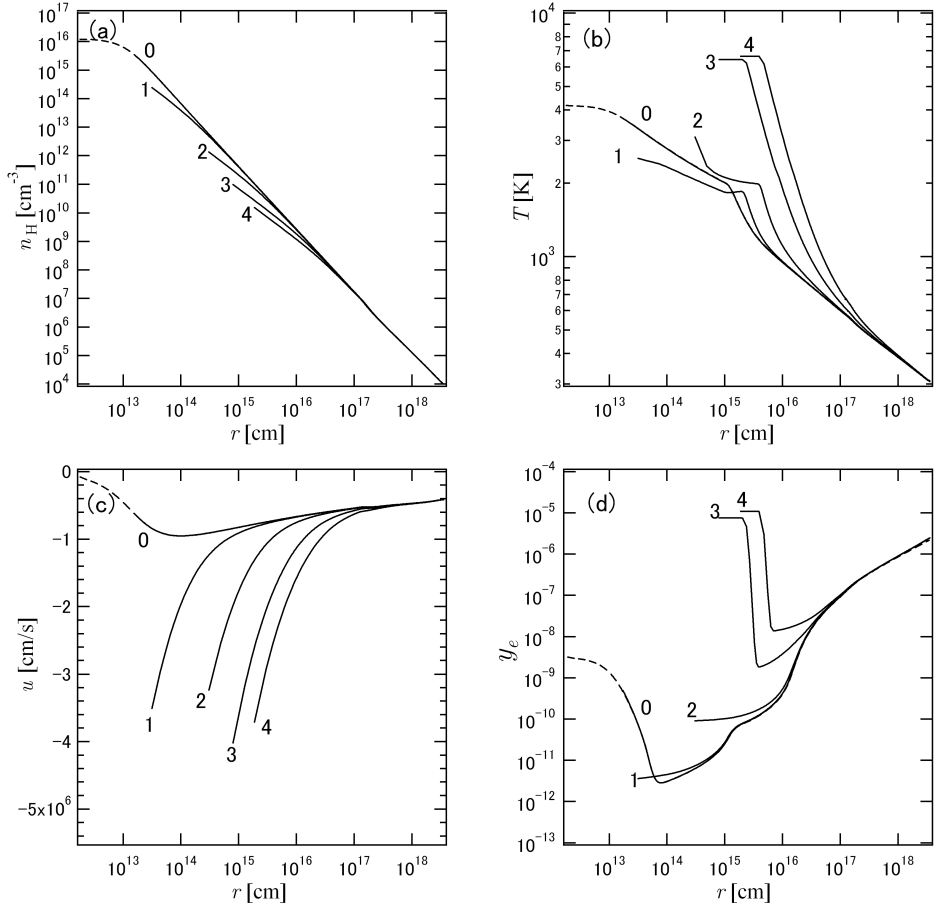}
\caption{The evolutionary sequences of the primordial gas cloud
in accretion phase are shown. Four panels (a)-(d) show the spatial
 distribution of following physical variables as functions of radius:
(a) number density $n_{\rm H}$,
(b) temperature $T$,
(c) velocity $u$, and
(d) electron fraction $y_{\rm e}$.
Five time sequences (0-4) are plotted. Accreted central mass 
$M_{\rm c}(t)$ is used as a clock. 
Corresponding mass at the stages are:~ 
0:$M_{\rm c}(t) = 8.87\times10^{-2}\;\MO$,
1:$M_{\rm c}(t) = 1.58\;\MO$,
2:$M_{\rm c}(t) = 12.7\;\MO$,
3:$M_{\rm c}(t) = 50.2\;\MO$,
4:$M_{\rm c}(t) = 100\;\MO$.
}
\label{fig:evol_sequences}
\end{center}
\end{figure}
\begin{figure}[tbh]
\begin{center}
\FigureFile(12cm,7cm){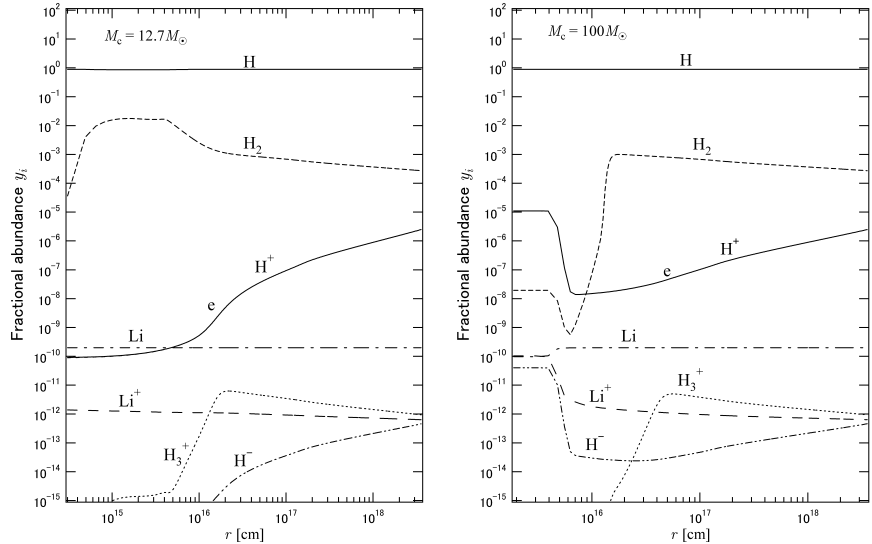}
\caption{The radial distributions of
the fractional abundances of the main species,
e, $\rm H^+$, H, $\rm H_2$, $\rm H^-$, $\rm H_3^+$, Li,
and $\rm Li^+$ in the accretion phase. Two panels correspond to the
 snapshots at $M_{\rm c}(t)=12.7\;\MO$(left), and $M_{\rm c}(t)=100\;\MO$(right).
}
\label{fig:chemi_accretion}
\end{center}
\end{figure}


\begin{figure}[htb]
\begin{center}
\FigureFile(12cm,7cm){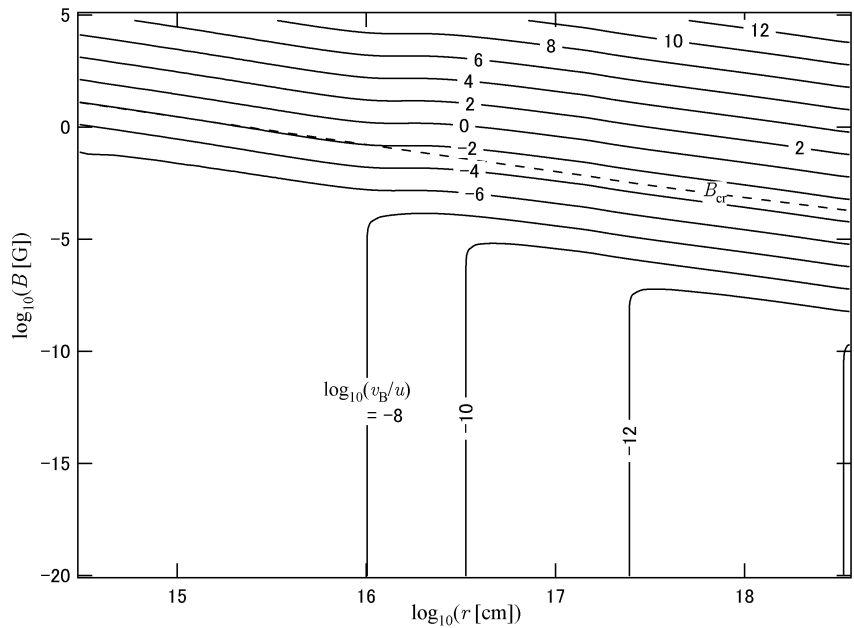}
\caption{The same as Fig.\ref{fig:rdiffu_runaway},
except that the contours are plotted in the accretion phase characterized
 at $M_{\rm c}$ =$12.7\;\MO$.}
\label{fig:rdiffu_accretion_127}
\end{center}
\end{figure}

\begin{figure}[htb]
\begin{center}
\FigureFile(12cm,7cm){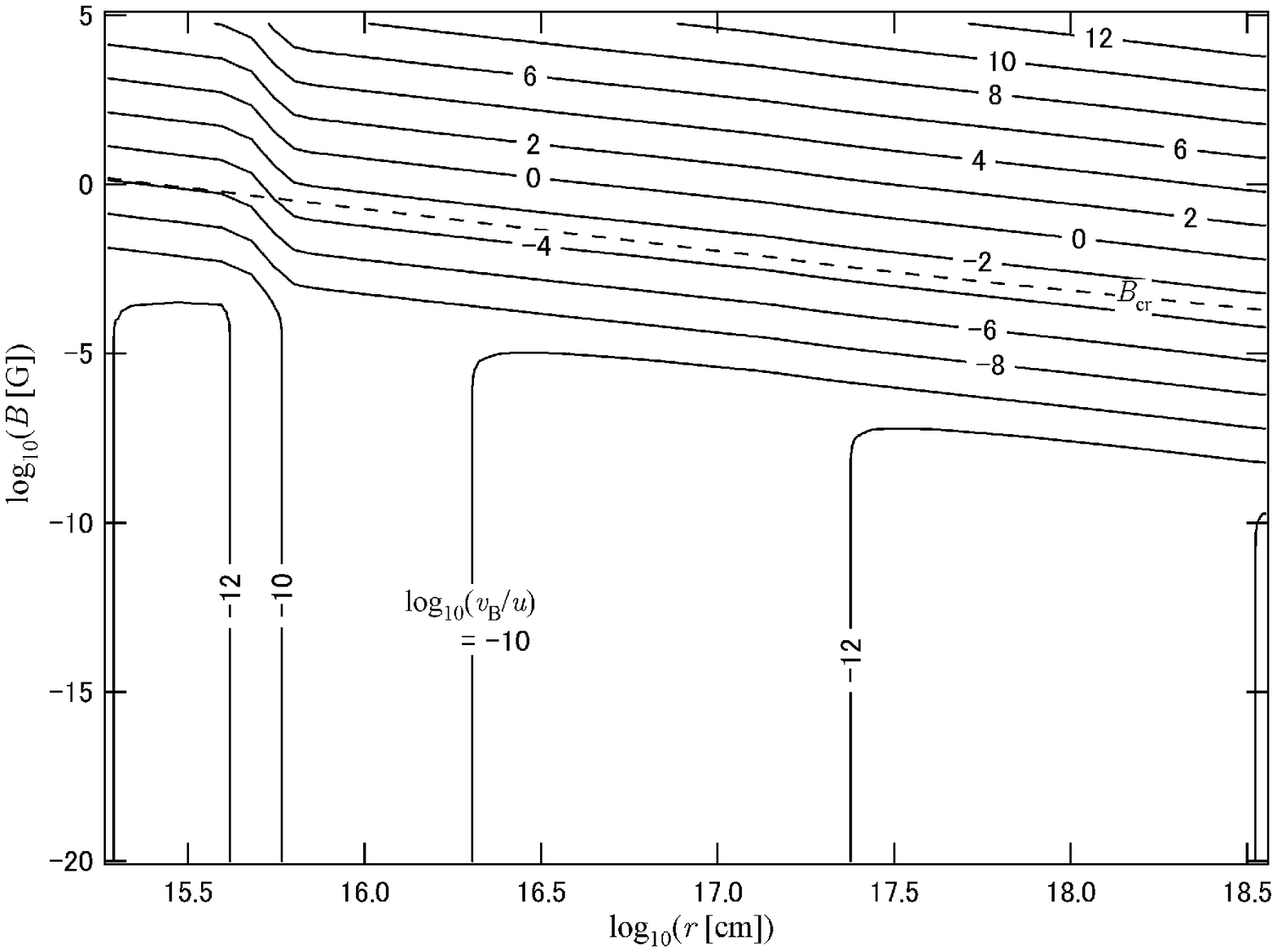}
\caption{The same as Fig.\ref{fig:rdiffu_accretion_127},
except $M_{\rm c}=100\;\MO$.}
\label{fig:rdiffu_accretion_100}
\end{center}
\end{figure}

\subsection{Accretion Phase}
\subsubsection{Evolution of gas and chemical species in accretion phase}

The evolutionary sequences of number density $n_{\rm H}$,
temperature $T$, velocity $u$, and
electron fraction $y_{\rm e}$ in the accretion phase
are illustrated in Figure \ref{fig:evol_sequences}.

Upper left panel illustrates the evolution of density profile. The solid
curves labeled as 0-4 corresponds to the epoch at which the central
accreted mass equals to $8.87\times10^{-2}\;\MO,1.58\;\MO,12.7\;\MO,50.2\;\MO,
100\;\MO$, respectively. 
It is clear that the density for $r\to 0$
decreases as the collapse proceeds, which is basically the same
feature found in similarity solution of polytropic cloud in accretion phase.

Because of the high accretion velocity at later snapshots (lower left), 
the gas is
heated efficiently by adiabatic compression. As a result, temperature
gets higher as the collapse proceeds, except at the beginning of the
accretion phase, when the gas is cooled efficiently due to the enhanced
fraction of H$_2$ molecules.

The ionization degree (electron fraction $y_{\rm e}$, lower right) also increases as
the collapse proceeds (Figures.\ref{fig:evol_sequences} and
\ref{fig:chemi_accretion}), due to the higher temperature for later snapshots.
As a result, $y_{\rm e}$ never gets lower than $10^{-11}$. We also find
that Li is not so important as was in the final phase of
run-away phase, since the hydrogen win back the position of chief
provider of electrons in the accretion phase( Figure \ref{fig:chemi_accretion}).

\subsubsection{Drift velocities in accretion phase}
As briefly discussed in section \ref{intro}, the ambipolar diffusion
velocity is proportional to the inverse square of gas density. 
Its dependence on various physical quantities are described as
$v_{{\rm B}x}^{\rm amb}
\propto y_{\rm e}^{-1}\rho^{-2}B^2r^{-1}$ (see equations (\ref{eq:diffusion_velocity})-(\ref{eq:A2})). 
On the other hand, the
accretion velocity scales as $u\propto (M_{\rm c}/r)^{1/2}$ for $r\to 0$.
Since we consider the accretion phase starting from the final phase of
run-away collapse with $\gamma=1.09$, the density profile and the
central mass  in accretion phase also depend on $\gamma$. 
According to \citet{suto88}, we have
\begin{eqnarray}
\rho(t,r)&\propto& t^{(2-3\gamma)/2}r^{-3/2}~~~~~~{\rm for~}r\to 0\\
M_{\rm c}(t)&\propto& t^{4-3\gamma}
\end{eqnarray}
Thus, the limiting behaviour of the infall velocity is described as
\begin{eqnarray}
u(t,r)\propto t^{(4-3\gamma)/2} r^{-1/2} ~~~~~~{\rm for~}r\to 0
\end{eqnarray}

Combining above set of equations, we have the dependence of the ratio
$v_{{\rm B}x}^{\rm amb}/u$ on $t$:
\begin{equation}
\frac{v_{{\rm B}x}^{\rm amb}}{u} \propto y_{\rm e}^{-1}B^2 t^{(9\gamma-8)/2}r^{5/2}
\end{equation}
This equation indicates that the ratio keeps growing in accretion phase 
for fixed $r,B$ and
$y_{\rm e}$, since $9\gamma -8$ is positive for $\gamma =1.09$.
In reality, however, $y_{\rm e}$ increases rapidly as the accretion
proceeds (see Figures.\ref{fig:evol_sequences} and
\ref{fig:chemi_accretion}), and it offsets the increase of $v_{{\rm B}x}^{\rm amb}/u$.

Figures 
\ref{fig:rdiffu_accretion_127} and \ref{fig:rdiffu_accretion_100}
illustrate
the contours of $\log_{10}(v_{{\rm B}x}/u)$ on $r-B$ plane for two
epochs. 
Two figures correspond to the snapshots
when the central accreted mass satisfies
$M_{\rm c} = 12.7\;\MO$ and $M_{\rm c} = 100\;\MO$, respectively.
The notations are same as Figure \ref{fig:rdiffu_runaway}.
It is clear that the drift velocity is always smaller than the accretion
velocity in two snapshots, as long as $B<B_{\rm cr}$.
In fact, we find that this is true all through the calculations. 
Besides, the equation (\ref{eq:B ratio}) holds also in the accretion phase.
Considering that the density profile in the accretion phase is $\rho\propto
r^{-1.5\sim -2}$,
$B$ is less than $B_{\rm cr}$ throughout the collapse, if ${B(t_0,r_0)}/{B_{\rm cr}(t_0,r_0)} < 1$, as discussed in \S\ref{drift_runaway}.
In other words, magnetic force is always negligible if it can be ignored
in the beginning of the collapse.

Thus, it is concluded that the dissipation of magnetic flux
is negligible throughout the mass accretion phase, as well as the
run-away collapse phase.
\section{DISCUSSION}
\label{discussion}
We confirmed that the magnetic field is frozen to the star-forming primordial
gas cloud even in the accretion phase. Here we discuss the possibility
of jet formation and activation of MRI considering the magnetic field
strength brought into the accretion disk surrounding the protostar.

The magnetic field strength brought into the accretion disk is assessed
under the frozen-in condition as follows:

\begin{equation}
B_{\rm disk} = B_0\left(\frac{R}{r_{\rm disk}}\right)^2.
\end{equation}
Here $B_0$ is the initial field strength, $r_{\rm disk}$ denotes the disk radius, whereas $R$ describes the
initial radius within which includes the
total mass of the disk-star star system
$M_{\rm *disk}$:
\begin{equation}
R = (3M_{\rm *disk}/4\pi \rho_0)^{1/3},
\end{equation}
where $\rho_0$ represents the initial density of the cloud.
$r_{\rm disk}$ can be evaluated by equation (15) in \citet{TM04}:
\begin{equation}
r_{\rm disk} \simeq 66.4 {\rm AU}\left(\frac{f_{\rm
		    Kep}}{0.5}\right)^2\left(\frac{M_{\rm *disk}}{10 \MO}\right)^{9/7}
\end{equation}
where $f_{\rm Kep}$ denotes the ratio of the rotation velocity to the
Kepler velocity of accreting matter, which is found to be $\sim 0.5$ in
\citet{abel02}. Thus, we have the magnetic field strength in the disk as
follows:

\begin{equation}
B_{\rm disk} \simeq 7.5\times 10^{-10} {\rm G}
\left(\frac{B_0}{3.7\times 10^{-16}{\rm G}}\right)
\left(\frac{n_{\rm H}}{10^3{\rm cm^{-3}}}\right)^{-2/3}
\left(\frac{f_{\rm Kep}}{0.5}\right)^{-4}
\left(\frac{M_{\rm *disk}}{10 \MO}\right)^{-40/21}
\label{eq:Bdisk}
\end{equation}

Several possibilities to generate cosmological seed magnetic field have
been proposed so far. Most of the mechanisms predict $B_{\rm IGM}\lesssim
10^{-19}$G\citep{widrow02}, except the magnetic field generated by
radiation transfer effects of powerful ionizing sources such as quasars
or first stars \citep{langer03}. They suggests the possibility to
generate coherent magnetic field with $B_{\rm IGM} \sim 10^{-11}$G.
Since the magnetic field is frozen to the primordial gas at low
densities ($n_{\rm H} < 10^3{\rm cm^{-3}}$), $B_0$ at $n_{\rm H} =
10^3{\rm cm^{-3}}$ can be evaluated as
\begin{equation}
B_0 = 3.7\times 10^{-16} {\rm G}
\left(\frac{B_{\rm IGM}}{10^{-19}{\rm G}}\right)
\left(\frac{n_{\rm H}}{10^3{\rm cm^{-3}}}\right)^{2/3}
\left(\frac{1+z}{20}\right)^2,\label{eq:B0}
\end{equation}

Recently, 3-dimensional MHD simulations on primordial star formation
have been performed by \citet{machida06}, assuming ideal MHD condition is
always satisfied in the collapsing gas. In fact, our present results
guarantee this hypothesis.
They found that 
the protostellar jet is driven in primordial environment
if $B_0\gtrsim 10^{-9}$G at $n_{\rm H} = 10^3{\rm cm^{-3}}$.
Comparing this condition with equation (\ref{eq:B0}), it is concluded
that 1) jets could be driven in first star forming clouds if the seed
field is generated by the mechanism proposed by \citet{langer03}, 2)
whereas the other mechanisms cannot generate the seed field enough to
drive the jets.

MRI can be activated in the accretion disk in case the magnetic field in
the disk is larger than a critical value \citep{TB04}:
\begin{equation}
	B_{\rm disk} \gtrsim 1.1\times 10^{-4}{\rm G} \left(\frac{M_{\rm *disk}}{10\ M_\odot}\right)^{1/4}
	\left(\frac{T}{10^4\ {\rm K}}\right)^{-3/4}
	\left(\frac{\ln\Lambda}{10}\right)^{1/2}
        \left(\frac{\rho_{\rm disk}}{5\times 10^{-10}\ {\rm g\;cm^{-3}}}\right)^{1/2}
	\left(\frac{r}{600\ R_\odot}\right)^{-3/4},
\label{eq:BMRI}
\end{equation}
This threshold is assessed by the confrontation between the growth rate
of MRI and the ohmic dissipation rate. Combining equations
(\ref{eq:Bdisk})- (\ref{eq:BMRI}), 
we find that MRI is activated in case the seed field satisfy
\begin{eqnarray}
B_{\rm IGM} \gtrsim 1.5 \times 10^{-14}{\rm G}&~&
\left(\frac{1+z}{20}\right)^{-2}
\left(\frac{f_{\rm Kep}}{0.5}\right)^4
\left(\frac{M_{\rm *disk}}{10 \MO}\right)^{2.155} \nonumber \\
&\times&\left(\frac{T}{10^4\ {\rm K}}\right)^{-3/4}
\left(\frac{\ln\Lambda}{10}\right)^{1/2}
\left(\frac{\rho_{\rm disk}}{5\times 10^{-10}\ {\rm g\;cm^{-3}}}\right)^{1/2}
\left(\frac{r}{600\ R_\odot}\right)^{-3/4}
.
\end{eqnarray}
Therefore, MRI is driven only if the seed field generation mechanism by
the transfer effects of ionizing radiation works. 
Based upon these arguments, we
emphasize that the mechanism proposed by
\citet{langer03} should be scrutinized since their results still based
upon the argument of order-estimation.


\section{SUMMARY}
\label{summary}
In this paper, we investigate
the dissipation of magnetic flux in star-forming primordial gas cloud.
We solve non-equilibrium chemical reaction equations, coupled with
thermal and dynamical evolution of the collapsing cloud all through the
run-away phase as well as the mass accretion phase. Thus, we obtain the
detailed evolution of ionized fraction of the gas, which enables us to
assess the coupling between gas and magnetic field.

As a result, we find that the magnetic field is 
basically 
frozen to the gas anywhere in {\it collapsing } star-forming primordial clouds at any time.
Based upon this result, we find the cosmological seed magnetic
field generated by most of the mechanisms proposed so far is not
sufficient to form jets as well as to activate MRI in the star-forming
cloud. Only one mechanism proposed by \citet{langer03} is able to create
sufficient field strength.

\bigskip

We thank Kazu Omukai for stimulating discussions. Noriaki Shibazaki and
Ken Ohsuga are acknowledged for continuous encouragement. 
The analysis has been
made with computational facilities at Rikkyo University.
This work was supported in part by Ministry of Education, Culture,
Sports, Science, and Technology (MEXT), Grants-in-Aid, Specially
Promoted Research 16002003 and Young Scientists (B) 17740110.

\appendix
\section{Density distribution of free-falling matter}
\label{appendix_a}
In this Appendix, we derive the explicit formula 
for $\displaystyle\left(\frac{\partial r}{\partial r_0}\right)_t$ which
gives the density distribution of free-falling matter in equation (\ref{eq:density in accretion phase}).

Equation of motion (\ref{eq:eq. of motion in accretion phase}) is integrated as
\begin{equation}
\frac{1}{2}\dot{r}^2=\frac{GM_0}{r} - \mathcal{E}_0,
\quad \mathcal{E}_0=\frac{GM_0}{r_0} - \frac{1}{2}u_0^2,
\quad (\mathcal{E}_0>0),
\label{eq:energy integral}
\end{equation}
where $u_0$ is the velocity of a fluid element at $(t_0,r_0)$ where $r_0$
denotes the position of the element at some initial time $t_0$.
A solution of this equation is given as
\begin{eqnarray}
r(\alpha,\alpha_0) &=& \frac{GM_0}{2\mathcal{E}_0}(1-\cos\alpha), \\
t(\alpha,\alpha_0) &=& t_0 - \frac{GM_0}{(2\mathcal{E}_0)^{3/2}}
    \left[(\alpha - \sin\alpha) - (\alpha_0 - \sin\alpha_0)\right],
\end{eqnarray}
where $\alpha$ is the so-called development angle, and $\alpha_0$
is its value at $t=t_0$. $\alpha_0$ also satisfies following relation
\begin{equation}
r_0 = \frac{GM_0}{2\mathcal{E}_0}(1-\cos\alpha_0).
\end{equation}

Using above relations, $(\partial r/\partial r_0)_t$ is derived as
\begin{equation}
\left(\frac{\partial r}{\partial r_0}\right)_t
 = \left(\frac{\partial r}{\partial M_0}\right)_{\mathcal{E}_0,\alpha}
   \left(\frac{\partial M_0}{\partial r_0}\right)_t
 + \left(\frac{\partial r}{\partial \mathcal{E}_0}\right)_{\alpha,M_0}
   \left(\frac{\partial \mathcal{E}_0}{\partial r_0}\right)_t
 + \left(\frac{\partial r}{\partial \alpha}\right)_{M_0,\mathcal{E}_0}
   \left(\frac{\partial \alpha}{\partial r_0}\right)_t,
\end{equation}
where
\begin{eqnarray}
\left(\frac{\partial r}{\partial M_0}\right)_{\mathcal{E}_0,\alpha}
 &=& \frac{G}{2\mathcal{E}_0}(1-\cos\alpha) = \frac{r}{M_0}, \\
\left(\frac{\partial r}{\partial \mathcal{E}_0}\right)_{\alpha,M_0}
 &=& -\frac{GM_0}{2\mathcal{E}_0^2}(1-\cos\alpha)=-\frac{r}{\mathcal{E}_0}, \\
\left(\frac{\partial r}{\partial \alpha}\right)_{M_0,\mathcal{E}_0}
 &=& \frac{GM_0}{2\mathcal{E}_0}\sin\alpha, \\
\frac{d\alpha_0}{dr_0}
 &=& \left(1 - \frac{r_0}{M_0}\frac{dM_0}{dr_0}
             + \frac{r_0}{\mathcal{E}_0}\frac{d\mathcal{E}_0}{dr_0}\right)
    /\left(\frac{GM_0}{2\mathcal{E}_0}\sin\alpha_0\right), \\
\left(\frac{\partial M_0}{\partial r_0}\right)_t
 &=& \frac{dM_0}{dr_0} = 4\pi r_0^2\rho_0, \\
\left(\frac{\partial \mathcal{E}_0}{\partial r_0}\right)_t
 &=& \frac{d\mathcal{E}_0}{dr_0}
  = \frac{G}{r_0}\frac{dM_0}{dr_0}-\frac{GM_0}{r_0^2}-u_0\frac{du_0}{dr_0}, \\
\left(\frac{\partial \alpha}{\partial r_0}\right)_t
 &=& \frac{r_0}{r}\frac{d\alpha_0}{dr_0}
    +\frac{\left(t-t_0\right)\sqrt{2\mathcal{E}_0}}{rM_0}\frac{dM_0}{dr_0}
    -\frac{3\left(t-t_0\right)}{r\sqrt{2\mathcal{E}_0}}\frac{d\mathcal{E}_0}{dr_0}.
\end{eqnarray}



\end{document}